\documentclass{revtex4-2}%
\usepackage[latin9]{inputenc}
\usepackage{amsmath}
\usepackage{amssymb}
\usepackage{graphicx}
\usepackage[unicode=true,  bookmarks=false,  breaklinks=false,pdfborder={0 0
1},backref=section,colorlinks=false]{hyperref}
\usepackage{amsfonts}
\usepackage{subfigure}
\usepackage{cleveref}
\usepackage{listings}
\usepackage{xcolor}
\usepackage{array}
\usepackage{url}%
\hypersetup{
colorlinks,linkcolor=blue,citecolor=blue,urlcolor=blue}
\makeatletter
\@ifundefined{textcolor}{}
{
\definecolor{BLACK}{gray}{0}
\definecolor{WHITE}{gray}{1}
\definecolor{RED}{rgb}{1,0,0}
\definecolor{GREEN}{rgb}{0,1,0}
\definecolor{BLUE}{rgb}{0,0,1}
\definecolor{CYAN}{cmyk}{1,0,0,0}
\definecolor{MAGENTA}{cmyk}{0,1,0,0}
\definecolor{YELLOW}{cmyk}{0,0,1,0}
}

\makeatother
\begin{document}
\preprint{CTP-SCU/2021011}
\title{Photon Spheres and Spherical Accretion Image of a Hairy Black Hole}
\author{Qingyu Gan}
\email{gqy@stu.scu.edu.cn}
\author{Peng Wang}
\email{pengw@scu.edu.cn}
\author{Houwen Wu}
\email{iverwu@scu.edu.cn}
\author{Haitang Yang}
\email{hyanga@scu.edu.cn}
\affiliation{Center for Theoretical Physics, College of Physics, Sichuan University,
Chengdu, 610064, China}

\begin{abstract}
In this paper, we first consider null geodesics of a class of charged,
spherical and asymptotically flat hairy black holes in an
Einstein-Maxwell-scalar theory with a non-minimal coupling for the scalar and
electromagnetic fields. Remarkably, we show that there are two unstable
circular orbits for a photon in a certain parameter regime, corresponding to
two unstable photon spheres of different sizes outside the event horizon. To
illustrate the optical appearance of photon spheres, we then consider a simple
spherical model of optically thin accretion on the hairy black hole, and
obtain the accretion image seen by a distant observer. In the single photon
sphere case, only one bright ring appears in the image, and is identified as
the edge of the black hole shadow. Whereas in the case with two photon
spheres, there can be two concentric bright rings of different radii in the
image, and the smaller one serves as the boundary of the shadow, whose radius
goes to zero at the critical charge.

\end{abstract}
\maketitle
\tableofcontents
\preprint{CTP-SCU/2021XXX}
\affiliation{Center for Theoretical Physics, College of Physics, Sichuan University,
Chengdu, 610064, PR China}

\section{Introduction}

The Event Horizon Telescope (EHT) collaboration has recently achieved an
angular resolution sufficient to observe the image of a supermassive black
hole in the center of galaxy M87, which allows us to test gravity in strong
field regime
\cite{Akiyama:2019cqa,Akiyama:2019brx,Akiyama:2019sww,Akiyama:2019bqs,Akiyama:2019fyp,Akiyama:2019eap}%
. The major feature of the image is a shadow region surrounded by a bright
ring, which results from strong gravitational lensing by the black hole
\cite{synge1966escape,bardeen1972rotating,bardeen1973timelike,Bozza:2009yw}.
The shadow image captured by EHT is expected to bear the fingerprint of the
geometry around the black hole, and in good agreement with the predictions of
the spacetime geometry of Kerr black holes. Nevertheless, the black hole
mass/distance and EHT systematic uncertainties still leave some room within
observational uncertainty bounds for a non-Kerr black hole. Moreover, there
has been considerable debate whether the bright ring surrounding the shadow is
solely determined by the photon sphere or also affected by details of the
accretion flow \cite{Gralla:2019xty,Narayan:2019imo}. So a lot of work in
progress has reported on black hole shadows for various black holes of
different theories of gravity with/without considering accretion flows
\cite{luminet1979image,Falcke:1999pj,Eiroa:2002mk,Yumoto:2012kz,Wei:2013kza,Zakharov:2014lqa,Atamurotov:2015xfa,Cunha:2016wzk,Dastan:2016bfy,Amir:2017slq,Wang:2017hjl,Ovgun:2018tua,Perlick:2018iye,Kumar:2019pjp,Zhu:2019ura,Ma:2019ybz,Mishra:2019trb,Zeng:2020dco,Zeng:2020vsj,Qin:2020xzu,Saurabh:2020zqg,Roy:2020dyy,Li:2020drn,Kumar:2020hgm,Zhang:2020xub,Mizuno:2018lxz,Peng:2020wun}%
.

On the other hand, the observation of a black hole shadow has become a new
venue to test the no-hair theorem
\cite{Israel:1967wq,Carter:1971zc,Ruffini:1971bza}, which states that a black
hole is uniquely characterized by its mass, angular momentum and electrical
charge. Various scenarios (e.g., black holes with Skyrme hairs
\cite{Luckock:1986tr,Droz:1991cx} and dilaton hairs \cite{Kanti:1995vq}, hairy
black holes in scalar-tensor gravities
\cite{Sotiriou:2013qea,Cisterna:2014nua} and Gauss-Bonnet theories
\cite{Antoniou:2017acq}) have been proposed to circumvent the no-hair theorem
since the first hairy black hole solution was found\ in\ the context of the
Einstein-Yang-Mills theory \cite{Volkov:1989fi,Bizon:1990sr,Greene:1992fw}.
For a review, see \cite{Herdeiro:2015waa}. Testing the no-hair theorem with
black hole shadows is crucial to understand black hole physics, and can be
used to constrain alternative theories of gravity. In light of this, studying
shadows of black holes with hair has attracted great attention
\cite{Johannsen:2010ru,Loeb:2013lfa,Cunha:2015yba,Cunha:2016bpi,Vincent:2016sjq,Cunha:2019ikd,moffat2020masses,Khodadi:2020jij}%
.

To gain better understanding of the formation of hairy black holes, an
Einstein-Maxwell-scalar (EMS) model with a non-minimal coupling of the scalar
field to the electromagnetic field has recently been put forward to study the
spontaneous scalarization in \cite{Herdeiro:2018wub}, where fully non-linear
numerical evolutions from scalar-free black holes to hairy black holes were
presented. Subsequently, many properties of this model and its extensions were
explored in the literature, e.g., various non-minimal coupling functions
\cite{Fernandes:2019rez,Blazquez-Salcedo:2020nhs}, dyons including magnetic
charges \cite{Astefanesei:2019pfq}, axionic-type couplings
\cite{Fernandes:2019kmh}, massive and self-interacting scalar fields
\cite{Zou:2019bpt,Fernandes:2020gay}, horizonless reflecting stars
\cite{Peng:2019cmm}, stability analysis of scalarized black holes
\cite{Myung:2018vug,Myung:2019oua,Zou:2020zxq,Myung:2020etf,Mai:2020sac},
higher dimensional scalar-tensor models \cite{Astefanesei:2020qxk},
quasinormal modes of scalarized black holes
\cite{Myung:2018jvi,Blazquez-Salcedo:2020jee}, two U(1) fields
\cite{Myung:2020dqt}, quasi-topological electromagnetism \cite{Myung:2020ctt},
topology and spacetime structure influences \cite{Guo:2020zqm}, the
Einstein-Born-Infeld-scalar theory \cite{Wang:2020ohb} and with a negative
cosmological constant \cite{Guo:2021zed,Zhang:2021etr}. In
\cite{Konoplya:2019goy}, the shadow radius in the EMS model was briefly
discussed to test the accuracy of the approximate analytical forms for the metric.

In this paper, we study the behavior of null geodesics (light rays) in the
hairy black hole solution obtained in \cite{Herdeiro:2018wub}, and use this to
obtain the image of a spherical accretion flow surrounding the black hole
perceived by a distant observer. A major new feature of photon motions in the
hairy black hole is that two unstable circular orbits for photons,
corresponding to two unstable photon spheres, exist outside the black hole
horizon in a certain parameter regime. It is found that the existence of two
unstable photon spheres affects the optical appearance of the accretion flow,
and gives rise to an additional bright ring and a smaller shadow. Note that
the intriguing feature of two unstable photon spheres has already been
reported in wormhole scenarios \cite{Shaikh:2018oul,Shaikh:2019jfr,Wielgus:2020uqz,Peng:2021osd,Tsukamoto:2021fpp,Guerrero:2021pxt}, yielding
some interesting optical phenomenons.

The rest of the paper is organized as follows. In Sec. \ref{sec:F}, we discuss
null geodesics in the black hole and describe the accretion model after the
hairy black hole solution is briefly reviewed. Section \ref{sec:PSS} contains
our main numerical results, which include effective potentials for photons,
trajectories of light rays and accretion images seen by a distant observer. We
conclude with a discussion in Sec. \ref{sec:Discussion-and-conclusion}. We set
$16\pi G=1$ throughout the paper.

\section{Framework}

\label{sec:F}

In this section, we review the hairy black hole solution, discuss features of
null geodesic motion, and introduce the accretion model. Specifically, we
consider the asymptotically flat black hole solutions with a non-trivial
scalar hair in an EMS theory with the exponential coupling
\cite{Herdeiro:2018wub},
\begin{equation}
S=\int d^{4}x\sqrt{-g}\left[  \mathcal{R}-2\partial_{\mu}\phi\partial^{\mu
}\phi-e^{\alpha\phi^{2}}F_{\mu\nu}F^{\mu\nu}\right]  , \label{eq:action}%
\end{equation}
where $\mathcal{R}$ is the Ricci scalar, the scalar field $\phi$ is minimally
coupled to the metric $g_{\mu\nu}$ and non-minimally coupled to the
electromagnetic field $A_{\mu}$, and $F_{\mu\nu}=\partial_{\mu}A_{\nu
}-\partial_{\nu}A_{\mu}$ is the electromagnetic tensor field. The action
$\left(  \ref{eq:action}\right)  $ has scalar-free black hole solutions with
the scalar field $\phi=0$, corresponding to Reissner-Nordstr\"{o}m (RN) black
holes. To obtain hairy black hole solutions, it showed that the dimensionless
coupling $\alpha$ has to be larger than $1/4$ \cite{Herdeiro:2018wub}.
Following \cite{Herdeiro:2018wub}, we focus on the static spherical black hole
solutions with the generic ansatz
\begin{equation}
ds^{2}=-N(r)e^{-2\delta(r)}dt^{2}+\frac{dr^{2}}{N(r)}+r^{2}\left(  d\theta
^{2}+\sin^{2}\theta d\varphi^{2}\right)  ,\qquad\mathbf{A}=A_{t}dt=V(r)dt.
\label{eq:metric ansatz}%
\end{equation}
The corresponding equations of motion are given by
\begin{align}
2m^{\prime}(r)-r^{2}N(r)\phi^{\prime}(r)^{2}-e^{2\delta(r)+\alpha\phi(r)^{2}%
}r^{2}V^{\prime}(r)^{2}  &  =0,\nonumber\\
\delta^{\prime}(r)+r\phi^{\prime}(r)^{2}  &  =0,\nonumber\\
\left[  e^{-\delta(r)}r^{2}N(r)\phi^{\prime}(r)\right]  ^{\prime}-\alpha
e^{\delta(r)+\alpha\phi(r)^{2}}\phi(r)r^{2}V^{\prime}(r)^{2}  &
=0,\label{eq:eom}\\
\left[  e^{\delta(r)+\alpha\phi(r)^{2}}r^{2}V^{\prime}(r)\right]  ^{\prime}
&  =0,\nonumber
\end{align}
where the Misner-Sharp mass function $m\left(  r\right)  $ is defined through
$N(r)\equiv1-2m(r)/r$, and the prime denotes the derivative with respect to
$r$. The last line in Eq. $\left(  \ref{eq:eom}\right)  $ leads to $V^{\prime
}(r)=-e^{-\delta(r)-\alpha\phi(r)^{2}}Q/r^{2}$, in which the constant $Q$ can
be interpreted as the electric charge of the black hole. Besides, we shall
implement suitable boundary conditions at the event horizon $r_{h}$,
\begin{equation}
m(r_{h})=\frac{r_{h}}{2},\delta(r_{h})=\delta_{0},\phi(r_{h})=\phi_{0}%
,V(r_{h})=0,
\end{equation}
and at the spatial infinity,
\begin{equation}
m(\infty)=M,\delta(\infty)=0,\phi(\infty)=0,V(\infty)=\Psi,
\end{equation}
where $\delta_{0}$ and $\phi_{0}$ are two constant parameters, $M$ is the ADM
mass, and $\Psi$ is the electrostatic potential. One can use the shooting
method to numerically solve the non-linear differential equations $\left(
\ref{eq:eom}\right)  $ for hairy black hole solutions satisfying the above
boundary conditions. Moreover, we focus on the fundamental state of hairy
black hole solutions, for which the scalar field $\phi(r)$ has none node.

To investigate the light deflection caused by a black hole, we need to find
the geodesic equation of light rays, which can be encapsulated in
\begin{equation}
\frac{d^{2}x^{\mu}}{d\lambda^{2}}+\Gamma_{\rho\sigma}^{\mu}\frac{dx^{\rho}%
}{d\lambda}\frac{dx^{\sigma}}{d\lambda}=0,\label{eq:geodesic}%
\end{equation}
with the affine parameter $\lambda$ and the Christoffel symbols $\Gamma
_{\rho\sigma}^{\mu}$. For the static spherical black hole $\left(
\ref{eq:metric ansatz}\right)  $, we can confine ourselves to light rays
traveling in the equatorial plane $\theta=\pi/2$, and introduce two conserved
quantities $E\equiv N(r)e^{-2\delta(r)}dt/d\lambda$ and $L\equiv r^{2}%
d\varphi/d\lambda$, which can be interpreted as the energy and the angular
momentum of the light rays, respectively. Therefore, the geodesic on the
equatorial plane for a light ray propagating in the metric $\left(
\ref{eq:metric ansatz}\right)  $ is given by
\begin{align}
\frac{dt}{d\eta} &  =\frac{1}{bN(r)e^{-2\delta(r)}},\label{eq:light ray eom1}%
\\
\frac{d\varphi}{d\eta} &  =\pm\frac{1}{r^{2}},\label{eq:light ray eom2}\\
\left(  \frac{dr}{d\eta}\right)  ^{2}+\frac{N(r)}{r^{2}} &  =\frac
{e^{2\delta(r)}}{b^{2}},\label{eq:light ray eom3}%
\end{align}
where the new affine parameter $\eta$ is related to the previous one by
$\eta=\lambda|L|$, the impact factor $b$ is defined as $|L|/E$, and $\pm$
correspond to moving in the counterclockwise $(+)$ and clockwise $(-)$ along
$\varphi$-direction, respectively. We can rewrite Eq. $\left(
\ref{eq:light ray eom3}\right)  $ as
\begin{equation}
e^{-2\delta(r)}\left(  \frac{dr}{d\eta}\right)  ^{2}+V_{\text{eff}}%
(r)=\frac{1}{b^{2}},
\end{equation}
where%
\begin{equation}
V_{\text{eff}}(r)=\frac{e^{-2\delta(r)}N(r)}{r^{2}}\label{eq:Veff}%
\end{equation}
is the effective potential. The trajectory of a light ray in the $r$-$\varphi$
plane is obtained by expressing $\varphi$ in terms of $r$ through Eqs.
$\left(  \ref{eq:light ray eom2}\right)  $ and $\left(
\ref{eq:light ray eom3}\right)  $, that is
\begin{equation}
\frac{d\varphi}{dr}=\pm\frac{1}{r^{2}e^{\delta(r)}\sqrt{\frac{1}{b^{2}%
}-V_{\text{eff}}(r)}}.\label{eq:phi-r}%
\end{equation}
Particularly, a circular null geodesic, dubbed photon sphere, occurs at
extrema of the effective potential $V_{\text{eff}}(r)$. In fact, the
conditions for the presence of a photon sphere are
\begin{equation}
V_{\text{eff}}(r_{ph})=\frac{1}{b_{ph}^{2}}\text{ and }V_{\text{eff}}^{\prime
}(r_{ph})=0,\label{eq:photon sphere}%
\end{equation}
where $r_{ph}$ is the radius of the photon sphere, and $b_{ph}$ is the
corresponding impact parameter. Moreover, maxima/minima of $V_{\text{eff}}(r)$
correspond to unstable/stable photon spheres. Since only unstable photon
spheres play an important role in determining properties of the accretion
image seen by a distant observer (e.g., the size of the black hole shadow), we
focus on unstable photon spheres in the following.

In this paper, we consider a toy model of a spherical illuminating accretion
flow, which is assumed to be optically transparent and statically distributed
outside the black hole horizon \cite{Narayan:2019imo}. This model is simple
but enough for the purpose of this paper. On the other hand, it is noteworthy
that M87 is known to contain a geometrically thick, optically thin, hot
accretion flow \cite{Yuan:2014gma}. The observed total photon intensity
$F_{o}$ (usually measured in ergs$^{-1}$cm$^{-2}$str$^{-1}$) at the celestial
point $\left(  X,Y\right)  $ in the observer's sky can be obtained by
integrating the emissivity along the photon path $\gamma$
\cite{Jaroszynski:1997bw,Kong:2013daa,Nampalliwar:2020asd,Bozza:2007gt,dexter2009fast}%
,
\begin{equation}
F_{o}\left(  X,Y\right)  =\int_{\nu_{o}}I_{o}\left(  \nu_{o},X,Y\right)
d\nu_{o}=\int_{\nu_{e}}\int_{\gamma}g^{4}j_{e}\left(  \nu_{e}\right)
dl_{\text{prop}}d\nu_{e}, \label{eq:Fo}%
\end{equation}
where $g$ is the redshift factor, and $I_{o}$ is the specific intensity at the
observed photon frequency $\nu_{o}$. Here, $\nu_{e}$, $dl_{\text{prop}}$ and
$j_{\text{e}}$ are the photon frequency, the infinitesimal proper length and
the emissivity per unit volume measured in the rest frame of the emitter,
respectively. For a photon of four-velocity $k^{\mu}=dx^{\mu}/d\eta$, we have
$g=k_{\alpha}u_{o}^{\alpha}/k_{\beta}u_{e}^{\beta}$ and $dl_{\text{prop}%
}=\left\vert k_{\beta}u_{e}^{\beta}d\eta\right\vert $, in which $u_{o}%
^{\alpha}$ and $u_{e}^{\alpha}$ are four-velocities of the distant observer
and the accretion emitter, respectively. In the model, we assume that the
illuminating accretion flow is at rest, which gives $u_{e}^{\alpha}%
=(e^{\delta(r)}/\sqrt{N(r)},0,0,0)$, and the distant observer is at the
spatial infinity, which gives $u_{o}^{\alpha}=\left(  1,0,0,0\right)  $. Using
Eqs. $\left(  \ref{eq:light ray eom1}\right)  $, $\left(
\ref{eq:light ray eom2}\right)  $ and $\left(  \ref{eq:light ray eom3}\right)
$, we then obtain%
\begin{equation}
g=\sqrt{N(r)}e^{-\delta(r)}\text{ and }dl_{\text{prop}}=\frac{e^{\delta(r)}%
dr}{b\sqrt{N(r)}\sqrt{\frac{e^{2\delta(r)}}{b^{2}}-\frac{N(r)}{r^{2}}}}.
\label{eq:g and dl}%
\end{equation}
Following
\cite{Jaroszynski:1997bw,Kong:2013daa,Nampalliwar:2020asd,Bozza:2007gt,dexter2009fast}%
, we also consider a simple case in which the specific emission is
monochromatic with the emitter's rest-frame frequency $\nu_{r}$, and has a
radial profile as
\begin{equation}
j_{e}(\nu_{e})=\frac{\delta(\nu_{r}-\nu_{e})}{r^{2}}. \label{eq:jeve}%
\end{equation}
Putting Eqs. $\left(  \ref{eq:g and dl}\right)  $ and $\left(  \ref{eq:jeve}%
\right)  $ into Eq. $\left(  \ref{eq:Fo}\right)  $, we find that the total
photon intensity measured by the distant observer can be expressed by
\begin{equation}
F_{o}(b)=\int_{\gamma}\frac{N(r)^{3/2}e^{-3\delta(r)}}{br^{2}\sqrt
{\frac{e^{2\delta(r)}}{b^{2}}-\frac{N(r)}{r^{2}}}}dr, \label{eq:F explicit}%
\end{equation}
with $b^{2}=X^{2}+Y^{2}$ due to the circular symmetry of the intensity.

\section{Photon Spheres and Shadows}

\label{sec:PSS}

\begin{figure}[tb]
\begin{centering}
\includegraphics[scale=0.56]{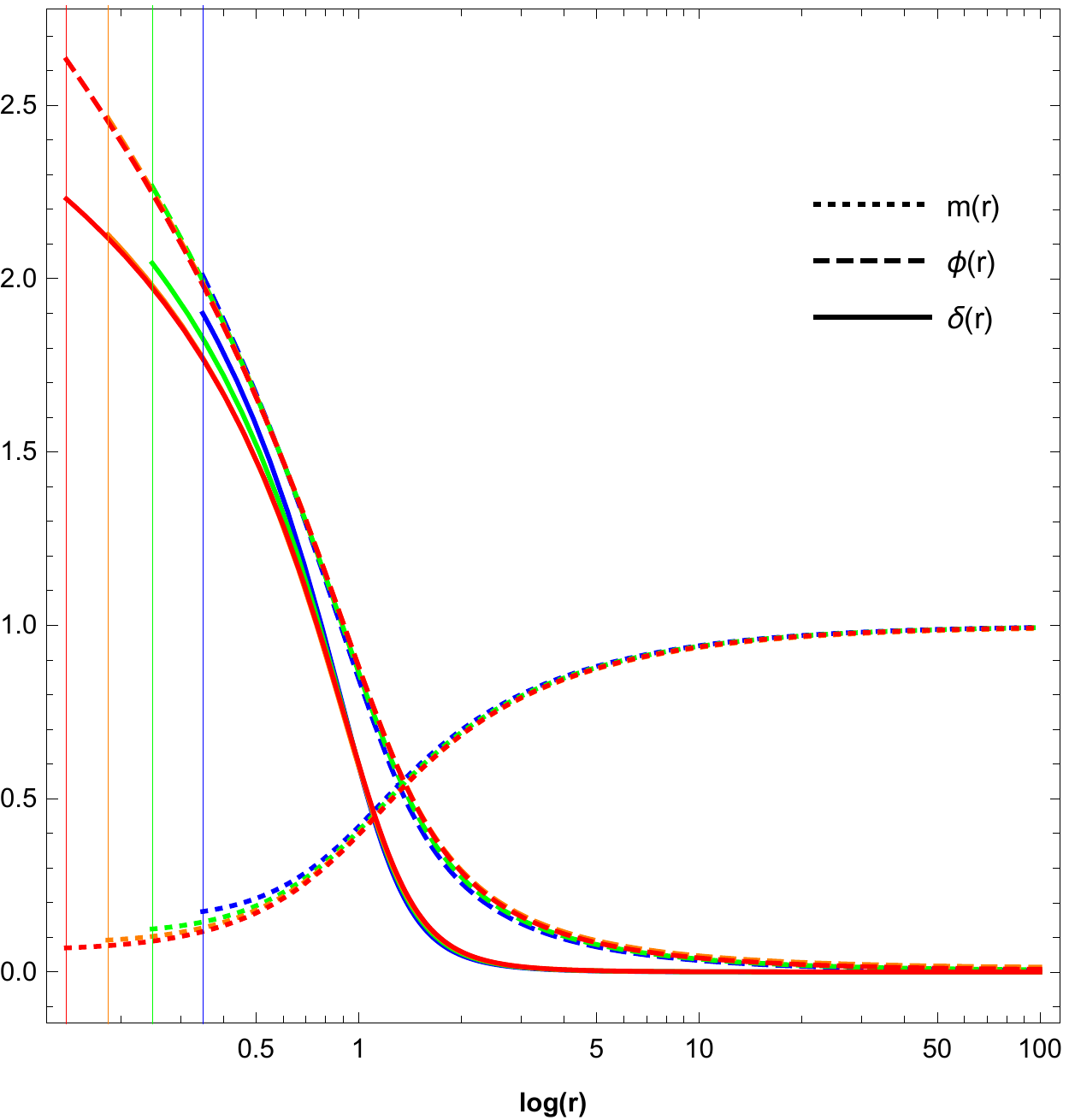}$\quad$\includegraphics[scale=0.6]{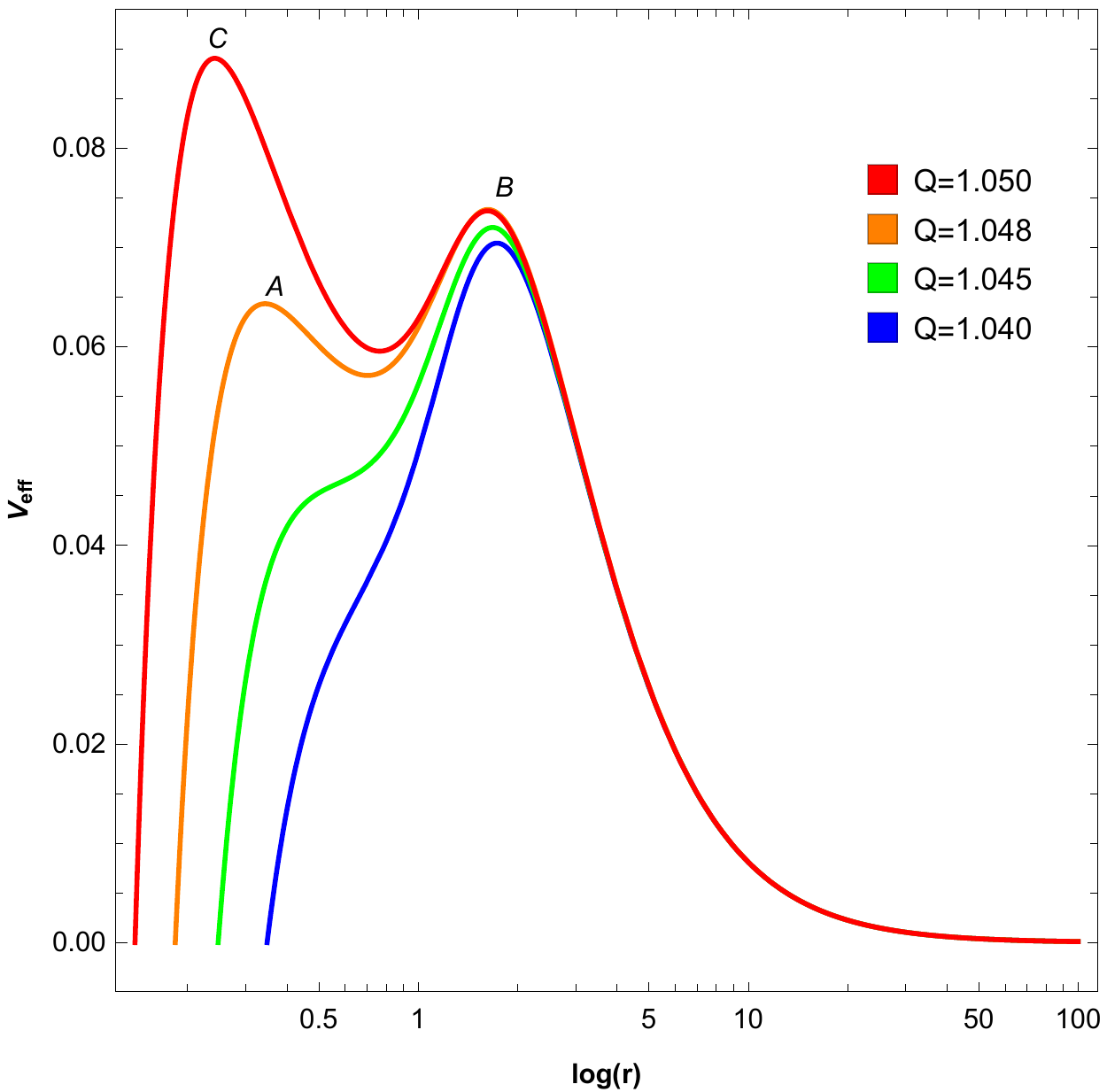}
\par\end{centering}
\centering{}\caption{Plots of hairy black hole solutions with $\alpha=0.8$ and
$M=1$ and the associated effective potentials $V_{\text{eff}}(r)$.
\textbf{Left}: Metric functions $m(r)$ (dotted), $\phi(r)$ (dashed) and
$\delta(r)$ (solid) for the hairy black holes with different electric charges
$Q=1.04$ (blue), $Q=1.045$ (green), $Q=1.048$ (orange) and $Q=1.05$ (red).
Vertical lines denote the corresponding event horizons. \textbf{Right}: For
$Q=1.04$ (blue) and $Q=1.045$ (green), the black hole effective potential
$V_{\text{eff}}(r)$ has a single-peak structure with a single maximum.
Nevertheless, a double-peak structure with two local maxima is observed for
$V_{\text{eff}}(r)$ when $Q=1.048$ (orange) and $Q=1.05$ (red). Note that the
global maximum of $V_{\text{eff}}(r)$ is responsible for determining the size
of the black hole shadow.}%
\label{figure Veff and sol}%
\end{figure}

In this section, we investigate the shadows and photon spheres of hairy black
hole solutions to the action $\left(  \ref{eq:action}\right)  $, which are
illuminated by static and spherical accretion flows. In the left panel of Fig.
\ref{figure Veff and sol}, we present the hairy black hole solutions for
several representative values of charge $Q$ with the coupling $\alpha=0.8$ and
the mass $M=1$. With $Q$ increasing, the size of event horizon of the hairy
black hole becomes smaller. The right panel of Fig. \ref{figure Veff and sol}
shows the corresponding effective potentials $V_{\text{eff}}(r)$, which have a
single maximum for $Q=1.04$ and $Q=1.045$, and two maxima for $Q=1.048$ and
$Q=1.05$. In what follows, the potentials with one maximum and two maxima are
dubbed single-peak and double-peak potentials, respectively. As discussed
before, the maxima of $V_{\text{eff}}(r)$ correspond to unstable photon
spheres, which can be responsible for determining the size of the black hole
shadow. Consequently, black hole solutions with single-peak (double-peak)
effective potentials possess one (two) unstable photon sphere(s). For
convenience, photon spheres refer to unstable photon spheres in the remainder
of this section. In the case with two photon spheres, the photon sphere with
smaller value of the impact parameter $b$ (e.g., the photon spheres $B$ and
$C$ for $Q=1.048$ and $1.05$, respectively, in the right panel of Fig.
\ref{figure Veff and sol}) determines the black hole shadow size
\cite{Junior:2021atr}.

\subsection{Single-peak potential}

\label{sec:One peak}

\begin{figure}[tb]
\begin{centering}
\includegraphics[scale=0.6]{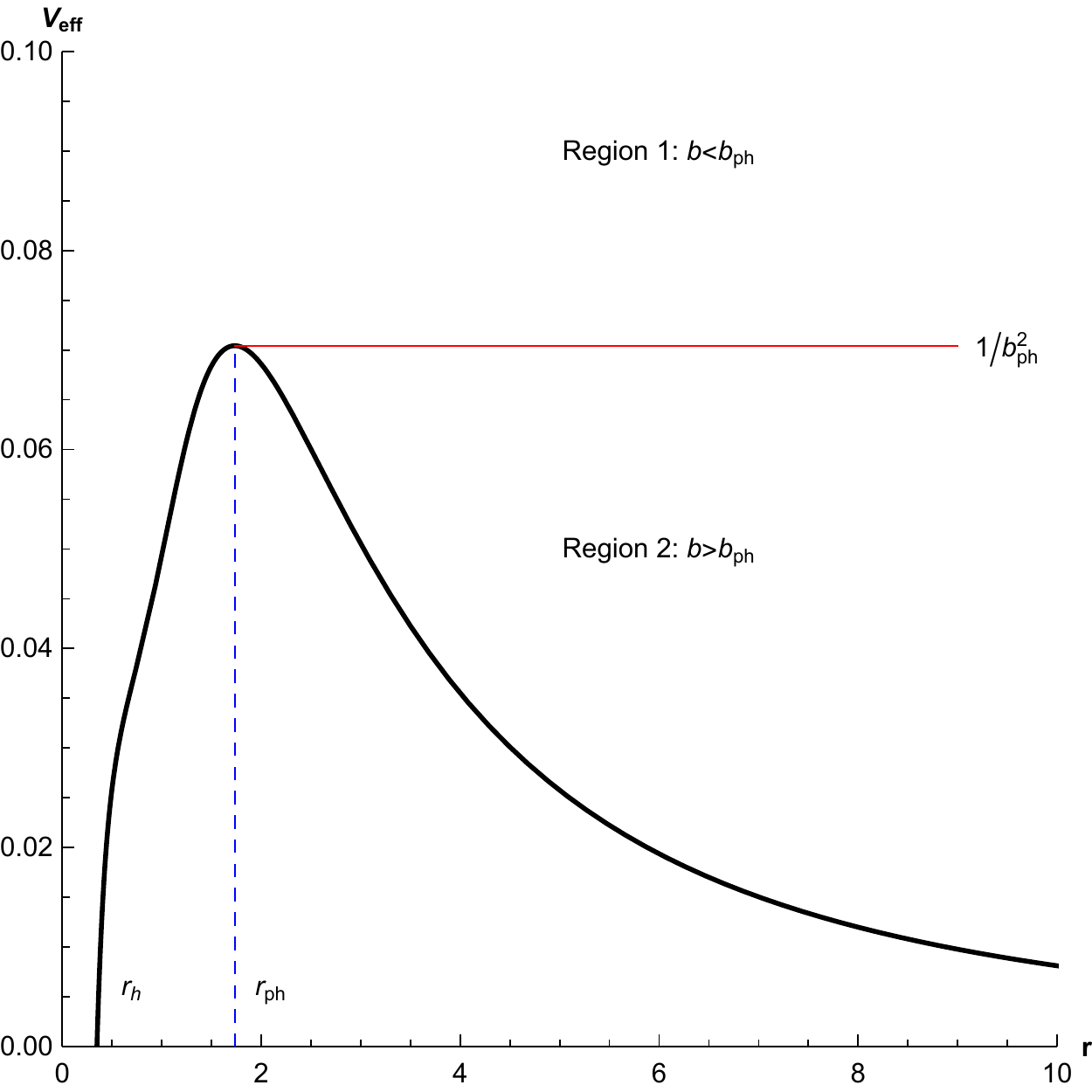}\includegraphics[scale=0.6]{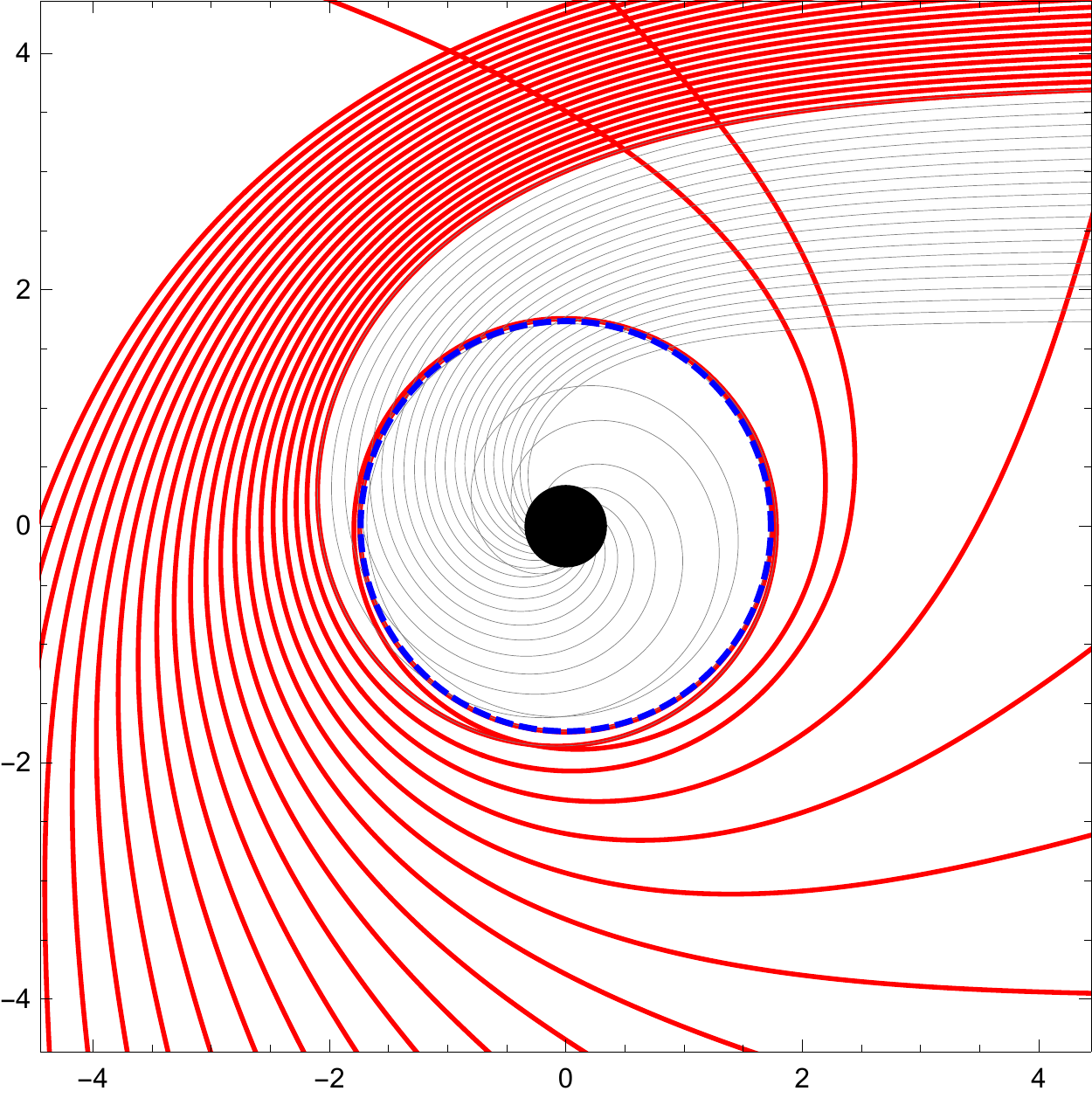}
\par\end{centering}
\caption{The profile of the effective potential $V_{\text{eff}}(r)$
(\textbf{Left}) and trajectories of light rays with different impact
parameters $b$ (\textbf{Right}) for the hairy black hole with $\alpha=0.8$,
$Q=1.04$ and $M=1$. The effective potential has a single maximum of
$1/b_{ph}^{2}$ at $r=r_{ph}$, where the photon sphere is located. The grey and
red lines correspond to light rays with $b<b_{ph}$ and $b>b_{ph}$,
respectively, and the dashed blue circle represents the photon sphere. The
black hole is shown as a solid black disk.}%
\label{figure single peak veff}%
\end{figure}

Here, we consider the shadow and photon sphere of the hairy black hole with
$\alpha=0.8$, $Q=1.04$ and $M=1$, which possesses a single-peak effective
potential. The effective potential and the trajectories of light rays are
plotted in Fig. \ref{figure single peak veff}. Suppose a distant observer is
located to the far right of the right panel of Fig.
\ref{figure single peak veff}. Thus, we consider a bundle of light rays
traveling towards the observer, which show different behavior depending on
their impact parameters $b$. In particular, light rays with $b=b_{ph}$
asymptotically approach the photon sphere of radius $r_{ph}$ and revolve
around a circular orbit at a constant radius $r=r_{ph}$ by infinite times. In
the parameter Region 1 with $b<$ $b_{ph}$, light rays (grey lines in the right
panel of Fig. \ref{figure single peak veff}) start from the black hole
horizon, and can overcome the barrier of the effective potential to propagate
to infinity. On the other hand, in the parameter Region 2 with $b>$ $b_{ph}$,
light rays (red lines in the right panel of Fig. \ref{figure single peak veff}%
) start from infinity, encounter the potential barrier at turning points, and
then reflect back in the outward direction towards the observer.

\begin{figure}[tb]
\begin{centering}
\includegraphics[scale=0.4]{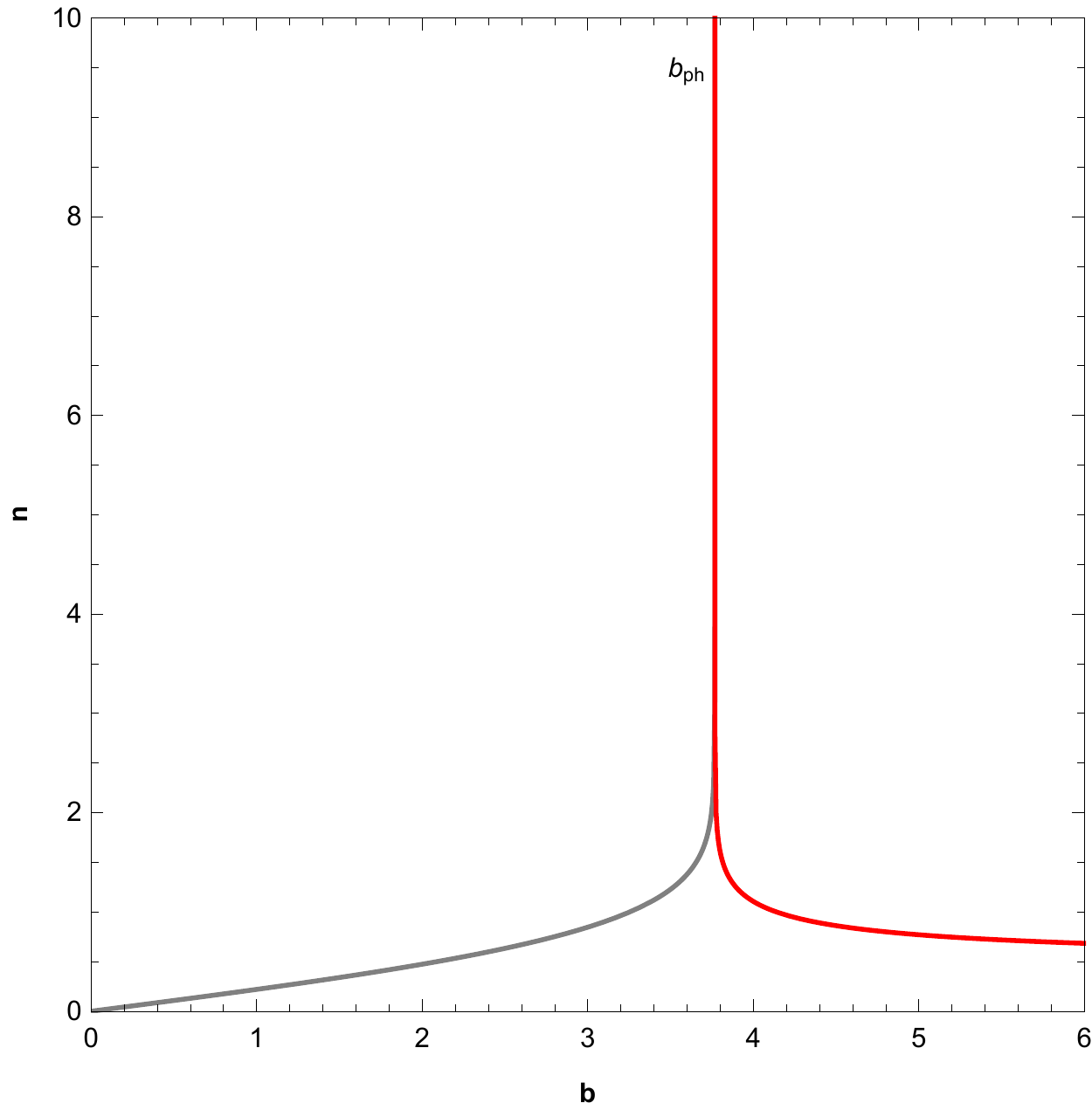}\includegraphics[scale=0.4]{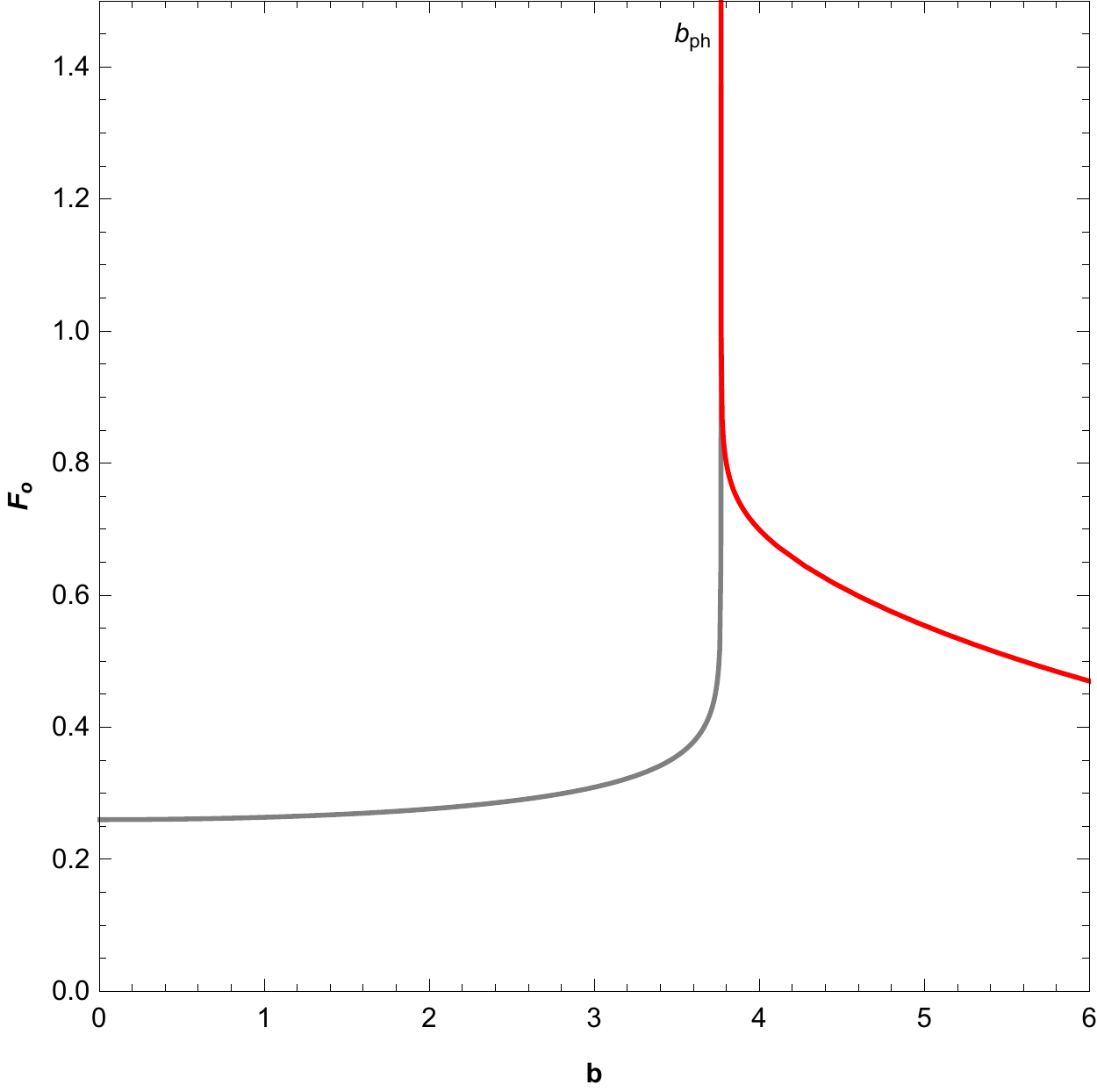}\includegraphics[scale=0.55]{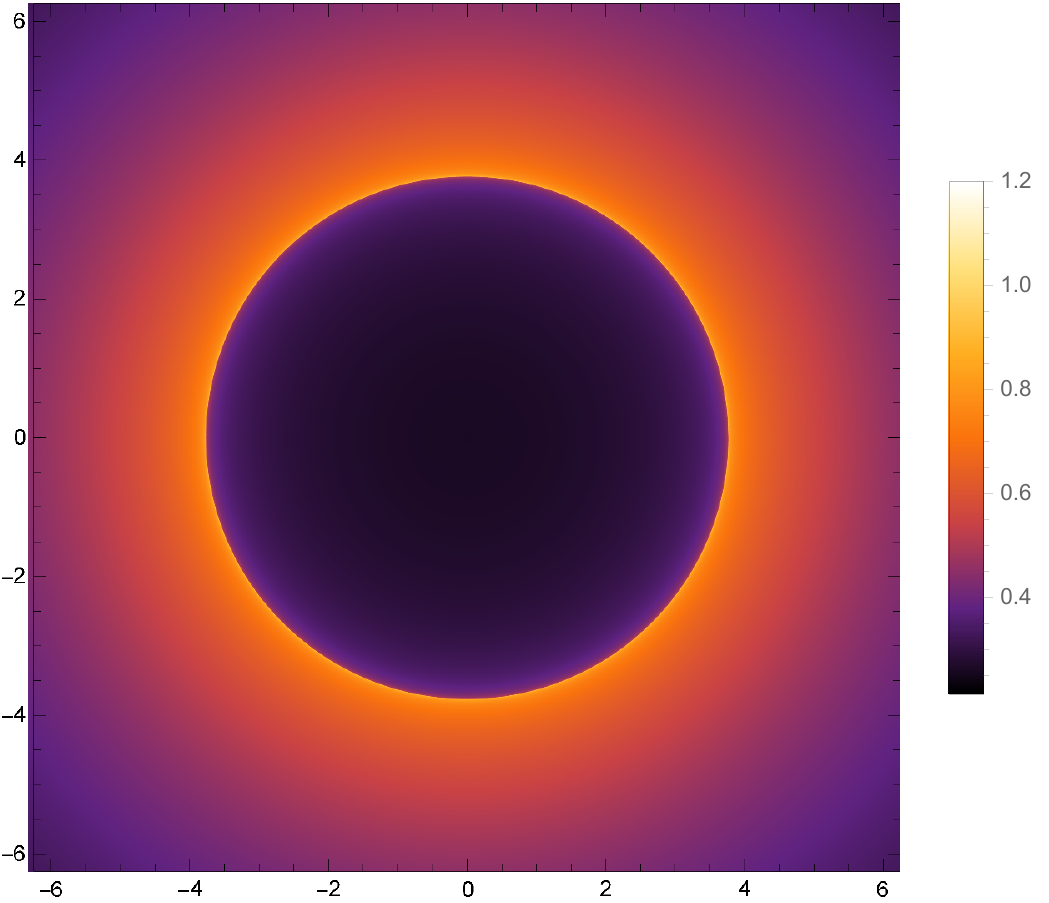}
\par\end{centering}
\caption{Plots of the total number of light ray orbits $n$, and the intensity
$F_{o}$ and accretion image seen by a distant observer for the hairy black
hole with $\alpha=0.8$, $Q=1.04$, and $M=1$. \textbf{Left}: The total number
of orbits $n\equiv\Phi/(2\pi)$ as a function of $b$, where $\Phi=\Delta
\varphi$ is the total change of the azimuthal angle of light rays traveling
outside the event horizon. The red (grey) segment is determined by light rays
with $b>b_{ph}$ ($b<b_{ph}$). \textbf{Middle}: The total photon intensity
$F_{o}(b)$ as a function of $b$, which has a sharp peak at $b=b_{ph}$. Light
rays with $b>b_{ph}$ ($b<b_{ph}$) contribute to the red (grey) segment.
\textbf{Right}: The 2D image of the accretion flow viewed in the observer's
sky. The bright ring is determined by the photon sphere, and serves as the
boundary of the black hole shadow.}%
\label{figure single peak intensity}%
\end{figure}

Using Eqs. $\left(  \ref{eq:phi-r}\right)  $ and $\left(  \ref{eq:F explicit}%
\right)  $, we plot the total number of light ray orbits $n$ and the observed
intensity $F_{o}$ as functions of $b$ in Fig.
\ref{figure single peak intensity}. As shown in the left panel, the light ray
with $b=b_{ph}$ orbits around the black hole an infinite number of times, and
hence picks up an arbitrarily large intensity from the accretion flow.
Consequently, the intensity observed by the distant observer rapidly increases
when $b$ increases towards $b_{ph}$, forms a sharp peak at $b_{ph}$, and then
decreases with increasing $b$, which is displayed in the middle panel. To
present the image of the accretion flow seen by the observer, we project
$F_{o}\left(  b\right)  $ to the observer's celestial coordinates $\left(
X,Y\right)  $ via $b^{2}=X^{2}+Y^{2}$ in the right panel. The 2D image has a
bright ring due to the peak of $F_{o}\left(  b\right)  $ at $b=b_{ph}$. The
dark region inside the bright ring refers to the black hole shadow, whose
intensity does not vanish since part of the radiation of the accretion flow
inside the photon sphere can escape to infinity. Note that this image is quite
similar to those given in
\cite{Zeng:2020dco,Zeng:2020vsj,Qin:2020xzu,Saurabh:2020zqg,Narayan:2019imo}.

\subsection{Double-peak potential}

\label{sec:double peak}

\begin{figure}[tb]
\begin{centering}
\includegraphics[scale=0.6]{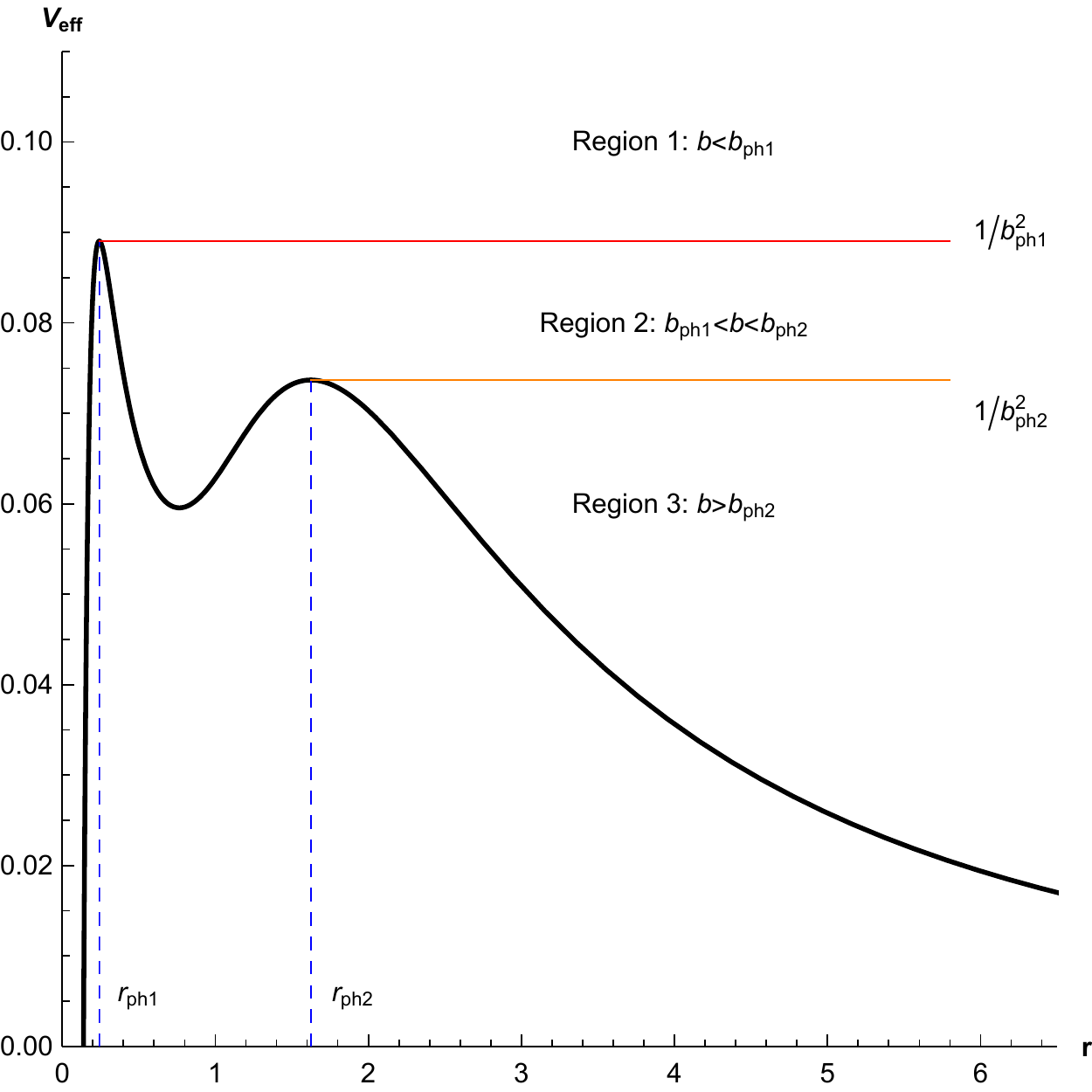}\includegraphics[scale=0.6]{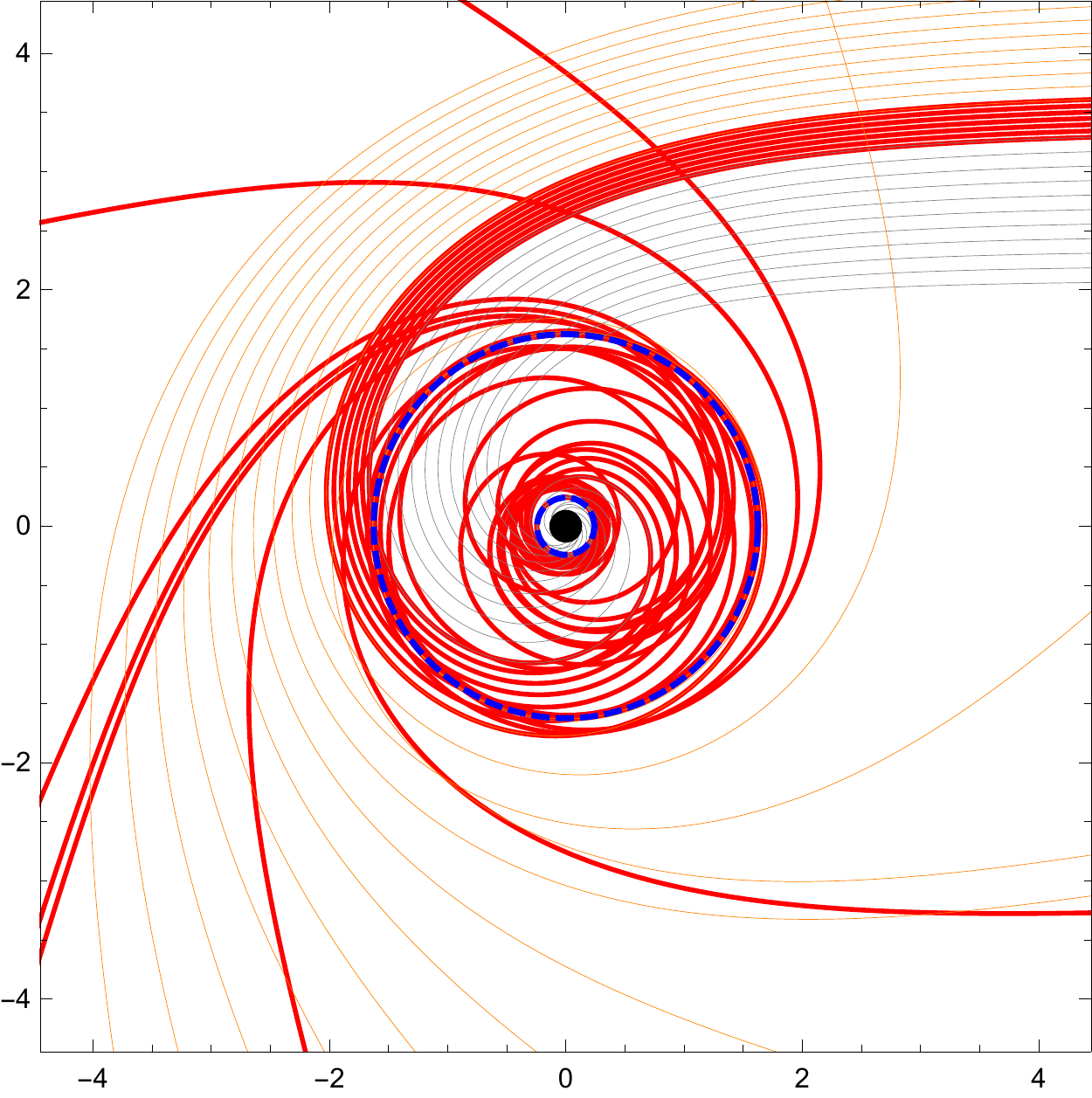}
\par\end{centering}
\begin{centering}
\includegraphics[scale=0.6]{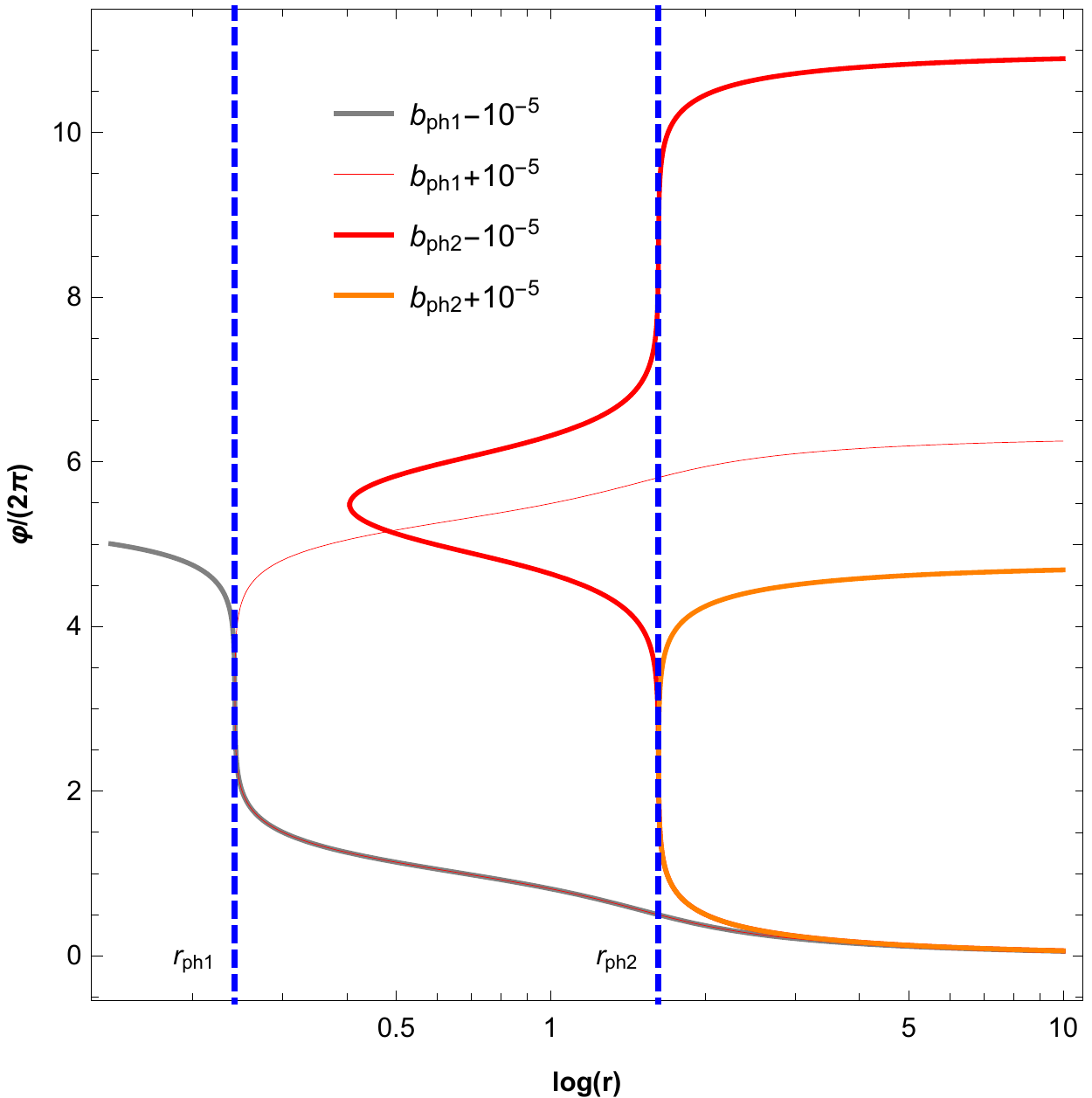}\includegraphics[scale=0.6]{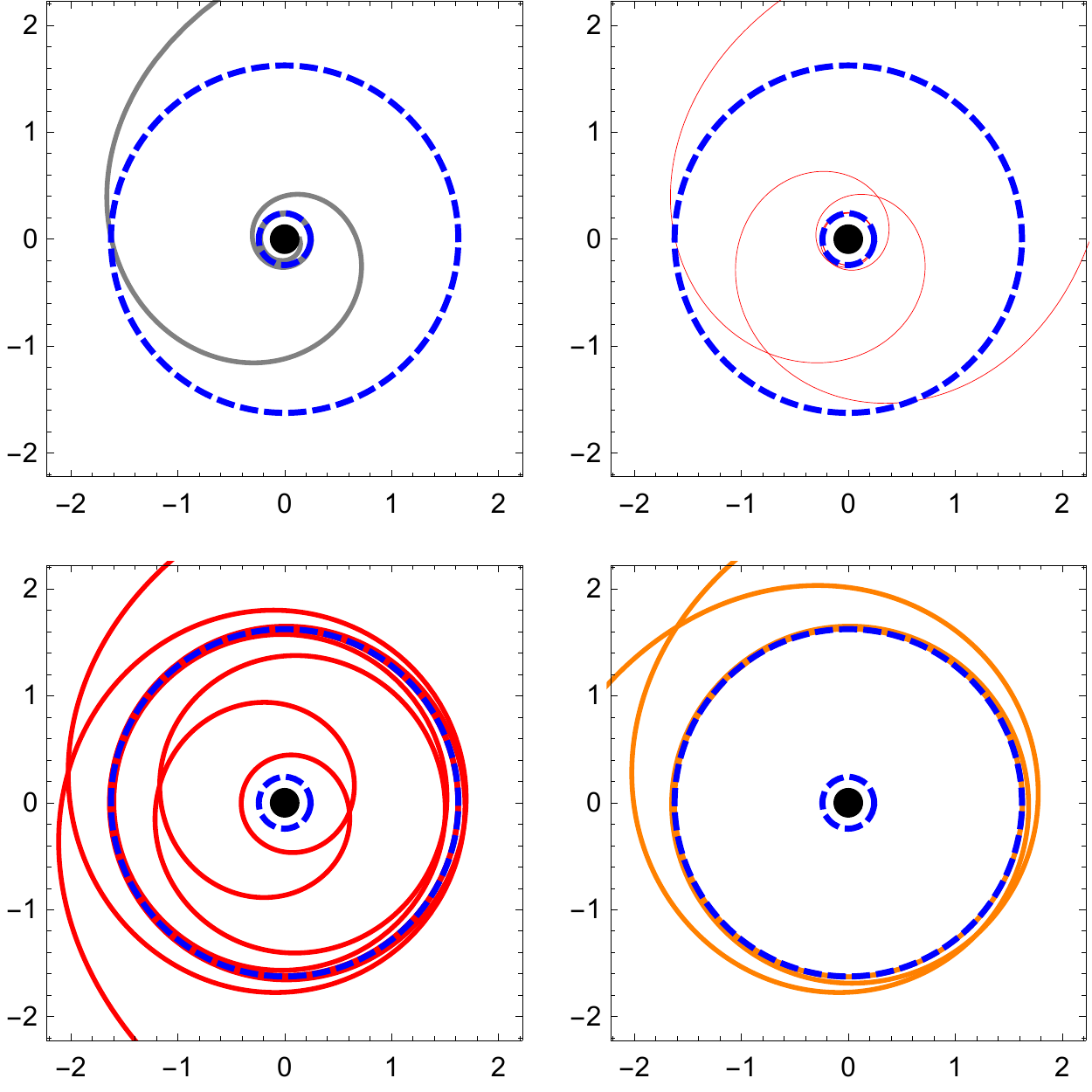}
\par\end{centering}
\caption{\textbf{Upper}: Plots of the effective potential and trajectories of
light rays for the hairy black hole with $\alpha=0.8$, $Q=1.04$ and $M=1$. The
effective potential has two maxima at $r=r_{ph1}$ and $r=r_{ph2}$,
corresponding to two photon spheres. Two dashed blue concentric circles of
radii $r_{ph1}$ and $r_{ph2}$ denote the two photon spheres. The grey, red and
orange lines represent light rays in Region 1, Region 2 and Region 3,
respectively. \textbf{Lower}: Plots of the fractional number of photon orbits
$\varphi(r)/(2\pi)$ and the corresponding trajectories for four light rays
with different $b$, which come from infinity and revolve around the photon
spheres many times before escaping. The dashed blue vertical lines in the left
panel correspond to the photon spheres.}%
\label{figure double peak veff}%
\end{figure}

A novel feature of the effective potential $V_{\text{eff}}(r)$ is that
$V_{\text{eff}}(r)$ can have two maxima for large enough black hole charge
$Q$, corresponding to two photon spheres outside the event horizon.
Nevertheless, these photon spheres do not always play a role in determining
the observed image of the accretion flow around a black hole. For instance,
considering the $Q=1.048$ case (orange) in Fig. \ref{figure Veff and sol},
light rays in the vicinity of the photon sphere associated with the peak $A$
can not escape to infinity since the effective potential $V_{\text{eff}}(r)$
at $B$ is higher than that at $A$, making this photon sphere invisible to the
distant observer. As a result, it is expected that the observed accretion
images of the $Q=1.04$, $Q=1.045$ and $Q=1.048$ cases in Fig.
\ref{figure Veff and sol} are quite similar. However, when the maximum of
$V_{\text{eff}}(r)$ occurring at a smaller $r$ is greater than that at a
larger $r$ (e.g., the $Q=1.05$ case in Fig. \ref{figure Veff and sol}), both
of the photon spheres are responsible for the accretion image seen by the
distant observer. In what follows, we focus on the hairy black hole with
$\alpha=0.8$, $Q=1.05$ and $M=1$ to study the effects of the two-peak
structure of $V_{\text{eff}}(r)$ on the observational appearance of the
accretion flow.

Using Eqs. $\left(  \ref{eq:Veff}\right)  $ and $\left(  \ref{eq:phi-r}%
\right)  $, we plot the corresponding effective potential and light ray
trajectories in the upper row in Fig. \ref{figure double peak veff}. Due to
the presence of two photon spheres, the behavior of light ray trajectories is
much richer than the single photon sphere case. As shown in the upper left
panel of Fig. \ref{figure double peak veff}, the effective potential features
two local maxima, $1/b_{ph1}^{2}$ and $1/b_{ph2}^{2}$, which occur at
$r=r_{ph1}$ and $r=r_{ph2}$, respectively. Since $b_{ph1}<b_{ph2}$ and
$r_{ph1}<r_{ph2}$, the parameter space of $b$ is divided into three regions,
in which trajectories of light rays behave differently. In Region 1 with
$b<b_{ph1}$, light rays coming from infinity go above the maximum of the
potential and get captured by the black hole. In Region 2 with $b_{ph1}%
<b<b_{ph2}$, light rays from infinity first travel toward the black hole, then
revolve around the smaller photon sphere until reaching the turning points,
and finally scatter off to infinity. In Region 3 with $b>b_{ph2}$, light rays
from infinity instead revolve around the larger photon sphere before escaping
to infinity. In Fig. \ref{figure double peak veff}, light rays in Region 1,
Region 2 and Region 3 are depicted by grey, red and orange lines, respectively.

To gain a better understanding of two photon spheres, we consider light rays
with impact parameters in the vicinity of $b_{ph1}$ and $b_{ph2}$, as shown in
the lower row of Fig. \ref{figure double peak veff}. For $b=b_{ph1}-10^{-5}$
and $b=b_{ph1}+10^{-5}$, both incident light rays move towards the black hole
and circle around the smaller photon sphere at $r_{ph1}$ by many times.
However, the light ray with $b=b_{ph1}-10^{-5}$ eventually falls into the
black hole, whereas that with $b=b_{ph1}+10^{-5}$ is reflected back to
infinity. The light rays with $b=b_{ph2}-10^{-5}$ and $b=b_{ph2}+10^{-5}$ both
revolve around the larger photon sphere at $r_{ph2}$ by many times before
escaping to infinity. Nevertheless, the former can go over the peak at
$r_{ph2}$ and approach the smaller photon sphere at $r_{ph1}$ while the latter
can not. Moreover, the closer $b$ of a light ray is to $b_{ph1}$ or $b_{ph2}$,
the more times it will revolve around the corresponding photon sphere. Thus it
is expected that light rays with $b=b_{ph1}$ or $b=b_{ph2}$ will
asymptotically approach the corresponding photon sphere and orbit around it by
infinite times.

\begin{figure}[tb]
\begin{centering}
\includegraphics[scale=0.4]{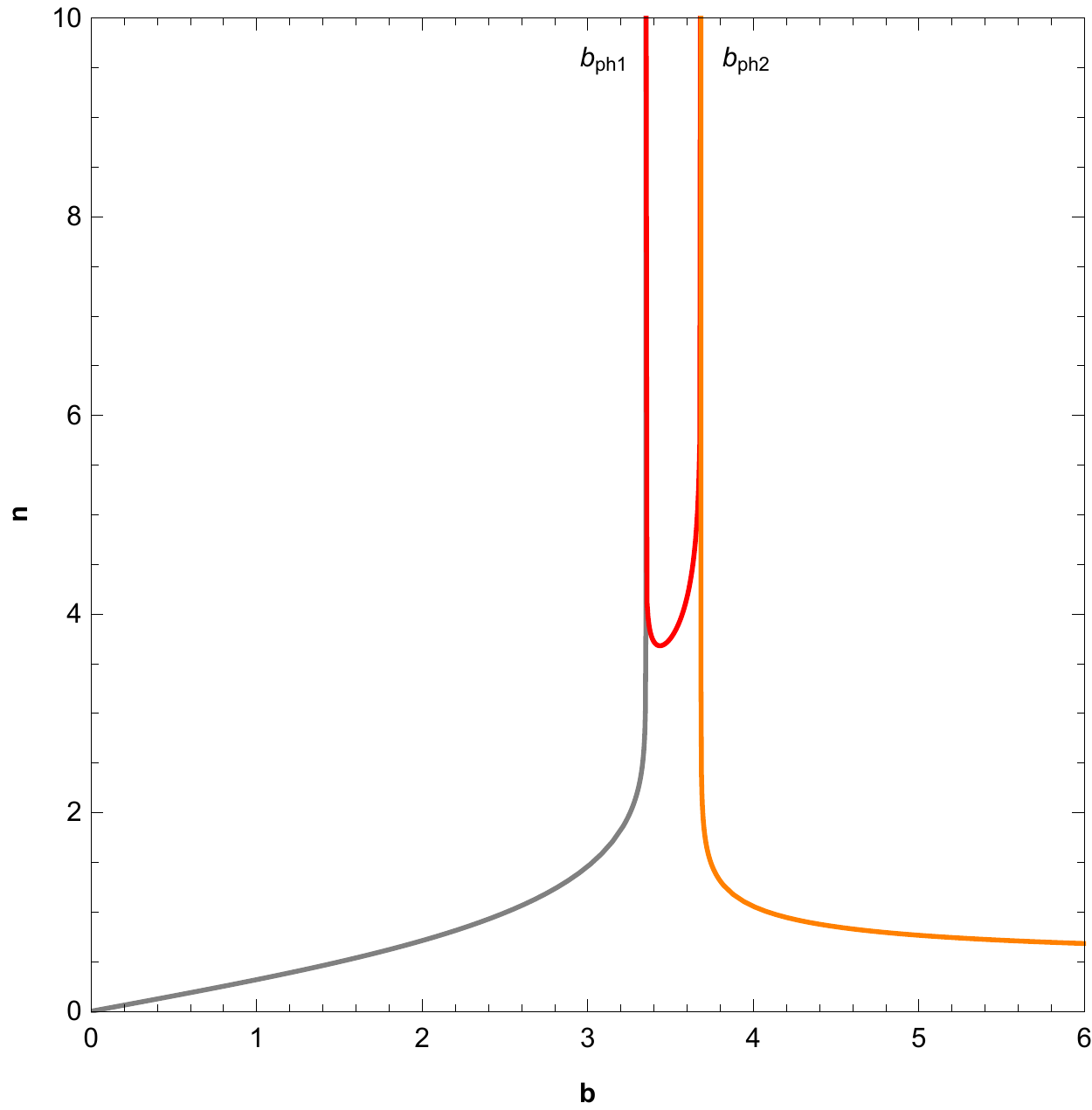}\includegraphics[scale=0.4]{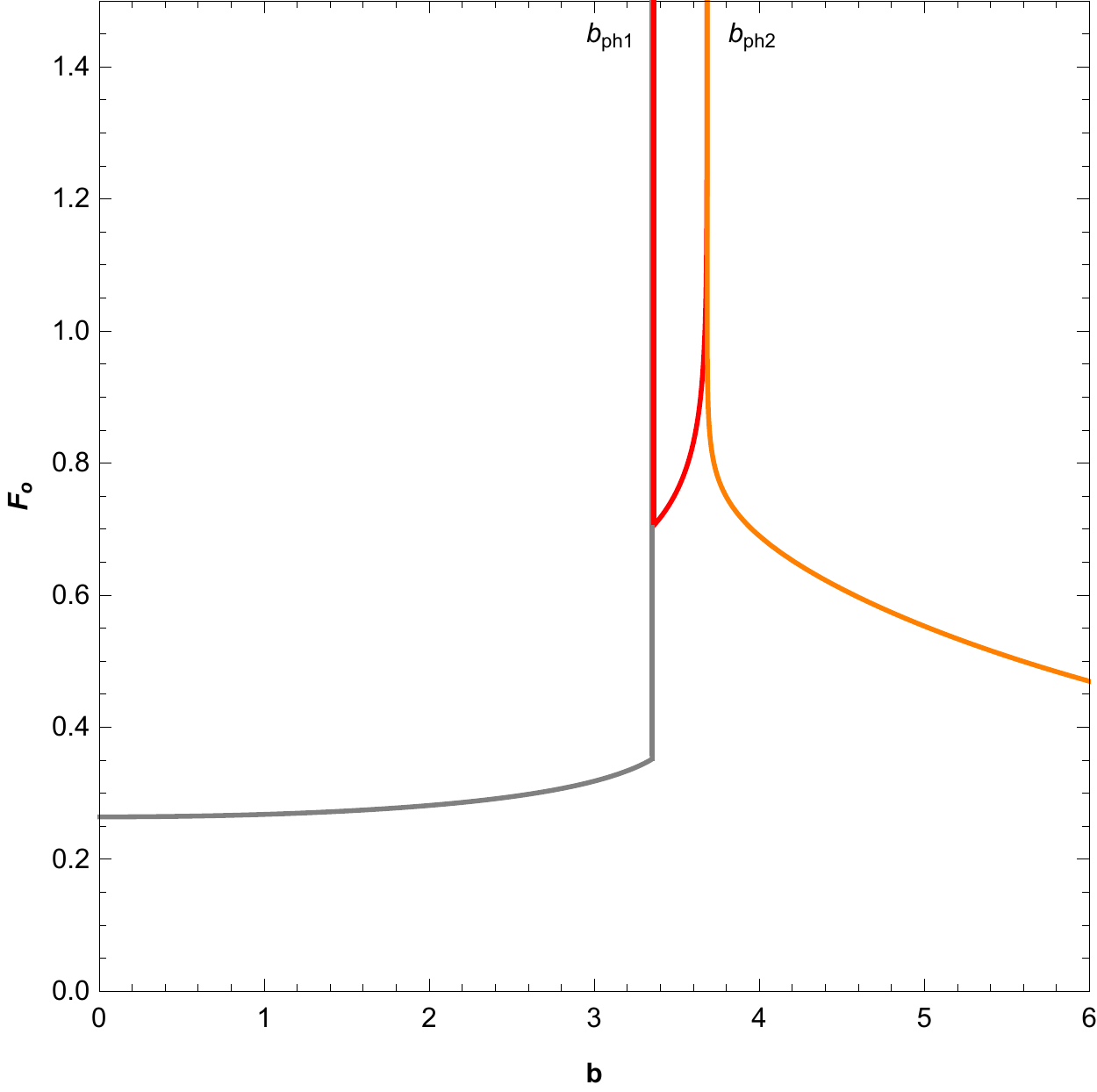}\includegraphics[scale=0.55]{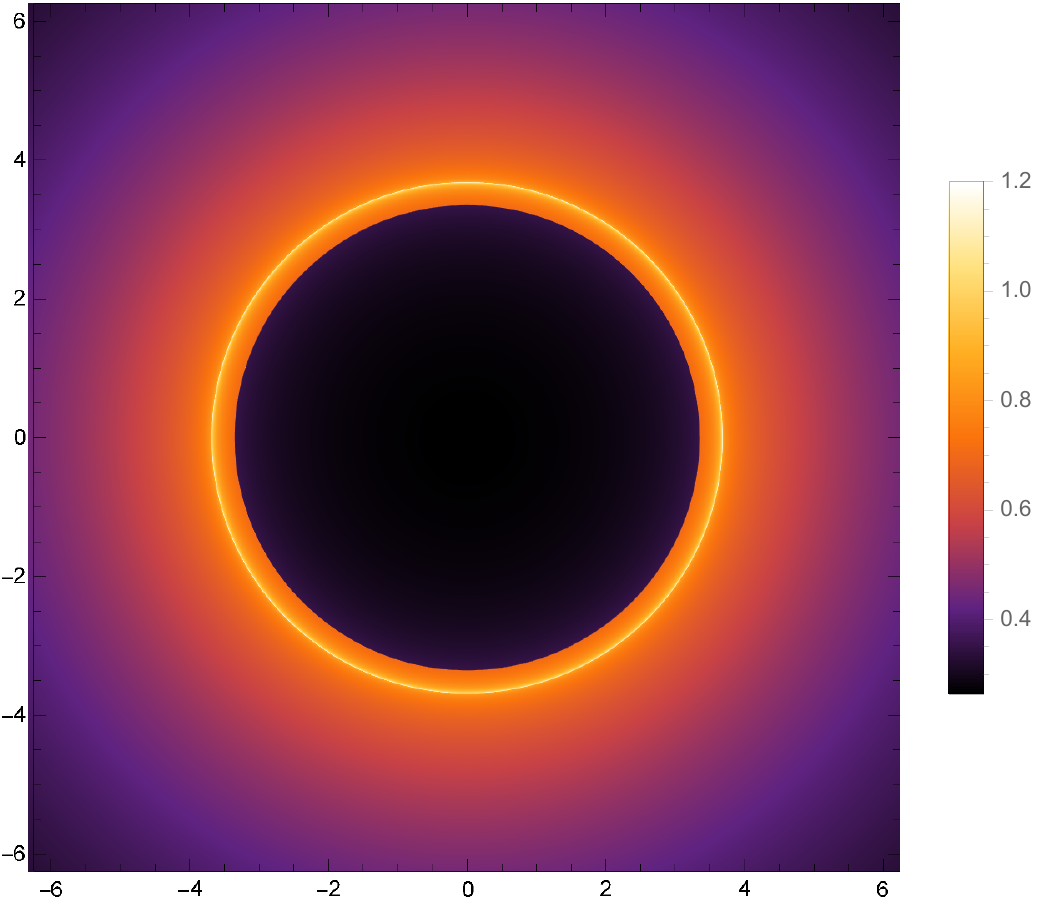}
\par\end{centering}
\caption{The total number of orbits $n\left(  b\right)  =\Phi/(2\pi)$
(\textbf{Left}), the observed total photon intensity $F_{o}(b)$
(\textbf{Middle}) and the observed 2D accretion image (\textbf{Right}) for the
hairy black hole with $\alpha=0.8$, $Q=1.05$ and $M=1$. The grey, red and
orange segments of $n\left(  b\right)  $ and $F_{o}(b)$ are determined by
light rays in Region 1, Region 2 and Region 3, respectively, in according with
colors used in Fig. \ref{figure double peak veff}. Due to the presence of two
photon spheres, $n\left(  b\right)  $ and $F_{o}(b)$ have two sharp peaks at
$b=b_{ph1}$ and $b=b_{ph2}$, corresponding to light rays that revolve around
the photon spheres at $r=r_{ph1}$ and $r=r_{ph2}$, respectively. Accordingly,
the accretion image viewed in the observer's sky has two concentric bright
rings of radii $b_{ph1}$ and $b_{ph2}$. The bright ring with the smaller
radius ($b_{ph1}$) is identified as the edge of the shadow, and barely visible
due to the narrow width of the peak of $F_{o}(b)$ at $b=b_{ph1}$.}%
\label{figure double peak intensity}%
\end{figure}

In Fig. \ref{figure double peak intensity}, we show the total number of light
ray orbits $n\left(  b\right)  $, the intensity profile $F_{o}\left(
b\right)  $ and the image of the static spherical accretion flow seen by a
distant observer in the hairy black hole with $\alpha=0.8$, $Q=1.05$ and
$M=1$. Unlike the single-peak case, there are two sharp peaks of the total
orbit number $n\left(  b\right)  $ at $b=b_{ph1}$ and $b=b_{ph2}$, which
result from the two photon spheres at $r=r_{ph1}$ and $r=r_{ph2}$.
Consequently, the observed intensity $F_{o}\left(  b\right)  $ has two peaks
at $b=b_{ph1}$ and $b=b_{ph2}$, which lead to two bright concentric rings of
the accretion image. As shown below, the bright ring with the larger radius
($b_{ph2}$), which is quite noticeable, is reminiscent of the bright ring of
radius $b_{ph}$ in the the accretion image in the single-peak case. However,
due to the sharpness of the peak of $F_{o}\left(  b\right)  $ at $b=b_{ph1}$,
the smaller bright ring, which locates at the edge of the shadow, is barely
visible. Note that, similar to the single-peak case, the intensity of the dark
shadow does not vanish completely due to the accretion radiation inside the
photon sphere.

\subsection{Dependence on black hole charge}

\begin{figure}[tb]
\begin{centering}
\includegraphics[scale=0.9]{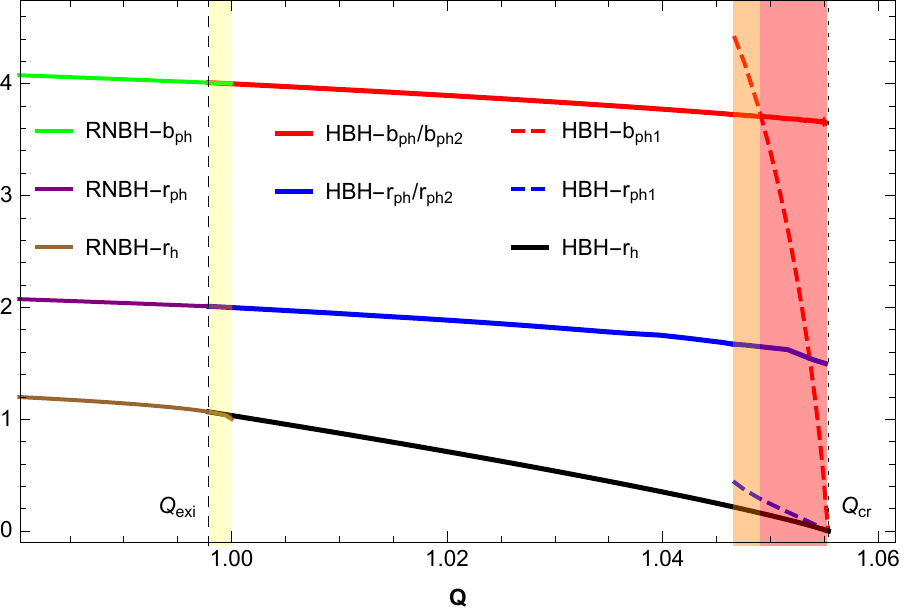}$\quad$\includegraphics[scale=0.65]{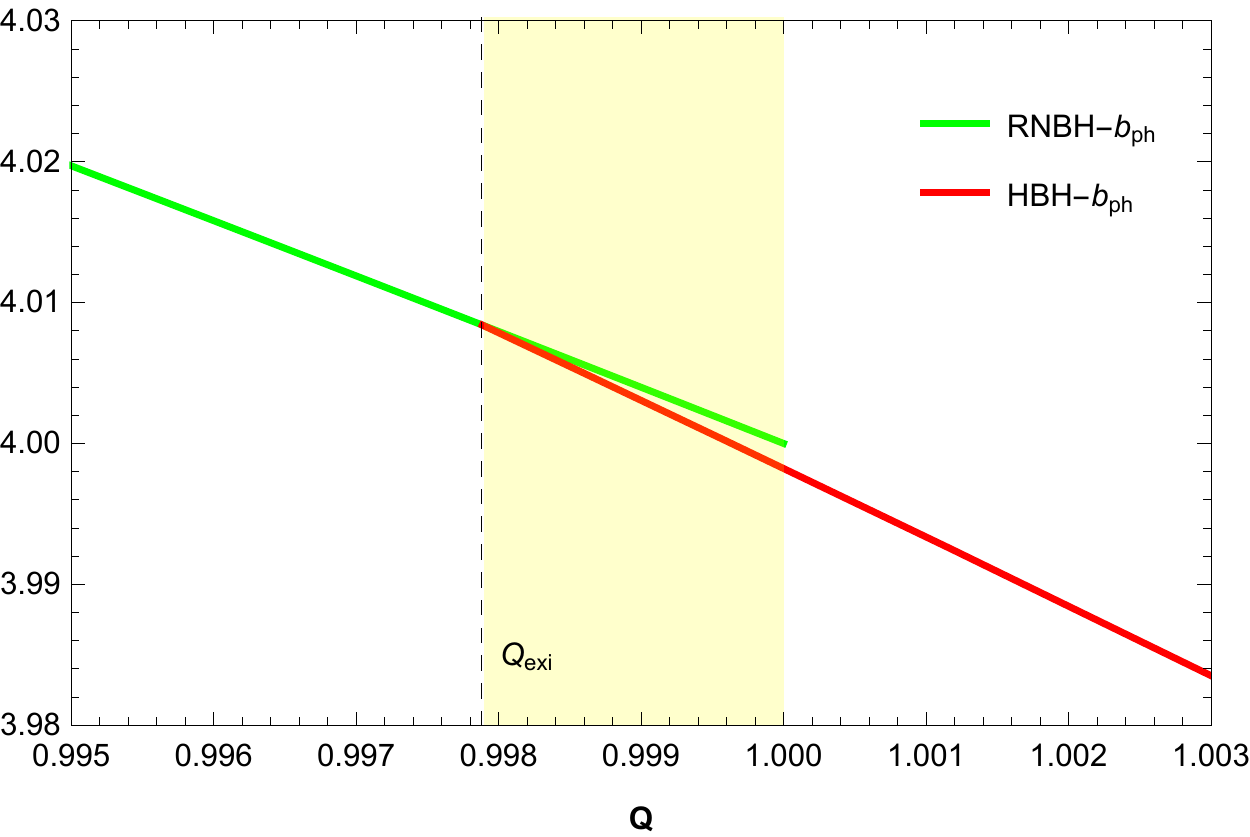}
\par\end{centering}
\centering{}\caption{Dependence of the photon sphere radii, $r_{ph}$,
$r_{ph1}$ and $r_{ph2}$, the associated impact parameters, $b_{ph}$, $b_{ph1}$
and $b_{ph1}$, and the horizon radius $r_{h}$ on the black hole charge $Q$ for
RN black holes with $M=1$ and hairy black holes with $\alpha=0.8$ and $M=1$.
The RN and hairy black holes coexist in the yellow region. In the orange and
pink regions, the hairy black holes have two photon spheres of radii $r_{ph1}$
and $r_{ph2}$ with $r_{ph1}<r_{ph2}$, and $b_{ph1}$ and $b_{ph2}$ are the
impact parameters associated with the corresponding photon spheres. The
minimum of $b_{ph1}$ and $b_{ph2}$ is identified as the radius of the black
hole shadow. In the orange region, $b_{ph2}<b_{ph1}$, and hence the shadow
radius is $b_{ph2}$, i.e., the impact parameter of the larger photon sphere.
Moreover, the smaller photon sphere of radius $r_{ph1}$ is not responsible for
imaging the accretion flow since it is invisible to the distant observer. In
the pink region, the shadow radius is the impact parameter $b_{ph1}$ of the
smaller photon sphere, and both photon spheres play a role in the observed
accretion image. The right panel highlights the impact parameters of the
photon spheres for the RN and hairy black holes in the coexisting region, and
shows that, for a given $Q$, the hairy black hole has a smaller shadow. }%
\label{fig:dependenceQ}%
\end{figure}

Here, we turn to the dependence of the size of the hairy black hole shadow on
the black hole charge. In Fig. \ref{fig:dependenceQ}, we plot the impact
parameters of light rays traveling at the photon spheres (red lines), the
radii of the photon spheres (blue lines) and the event horizon radius (black
line) as functions of the black hole charge for hairy black holes with
$\alpha=0.8$ and $M=1$. In \cite{Herdeiro:2018wub}, it showed that the hairy
black holes exist for some range of the black hole charge $Q$, $Q_{\text{exi}%
}\leq Q\leq Q_{\text{cr}}$. Here, $Q_{\text{exi}}$ is the existence charge,
for which a hairy hole solution bifurcates from a RN black hole solution, and
$Q_{\text{cr}}$ is the critical charge, for which the black hole horizon
radius vanishes with the black hole mass and charge remaining finite. In Fig.
\ref{fig:dependenceQ}, the vertical black dashed and dotted lines represent
$Q=Q_{\text{exi}}$ and $Q=Q_{\text{cr}}$, respectively. When $Q$ increases
from $Q_{\text{exi}}$, we find that the hairy black holes possess a single
photon sphere, and the left panel of Fig. \ref{fig:dependenceQ} shows that the
photon sphere radius $r_{ph}$ (solid blue line) and the impact parameter
$b_{ph}$ (solid red line) both decrease. For a large enough $Q~$(i.e., in the
orange and pink regions), apart from the photon sphere of radius $r_{ph1}$
(solid blue line), which is reminiscent of the photon sphere of radius
$r_{ph}$ in the single-peak case, there appears a new smaller photon sphere of
radius $r_{ph2}$ (dashed blue line), corresponding to the aforementioned
double-peak case. The size of the shadow is determined by the minimum of
$b_{ph1}$ (dashed red line) and $b_{ph2}$ (solid red line), which are the
impact parameters of the smaller and larger photon spheres, respectively.
Therefore, the shadow radii of the black holes in the orange and pink regions
are $b_{ph2}$ and $b_{ph1}$, respectively. The left panel shows that the black
hole shadow shrinks with increasing $Q$ and vanishes at $Q=Q_{\text{cr}}$,
where the black hole horizon becomes zero. Note that the shadow radius
decreases at a larger decreasing rate in the pink region.

For comparison, the photon sphere radius $r_{ph}$ (purple line), the
corresponding impact parameter $b_{ph}$ (green line) and the horizon radius
$r_{h}$ (brown line) of RN black holes with $M=1$ are displayed in Fig.
\ref{fig:dependenceQ}, which shows that $r_{ph}$, $b_{ph}$ and $r_{h}$ all
decrease when $Q$ increases. In the yellow region with $Q_{\text{exi}}\leq
Q\leq1$, the RN and hairy black holes coexist. As shown in the right panel of
Fig. \ref{fig:dependenceQ}, for a given $Q$ in the coexisting region, the
hairy black hole has a smaller impact parameter of the photon sphere than the
RN black hole. As a result, the shadow of the hairy black hole is smaller than
that of the RN black hole with the same charge.

\section{Discussion and Conclusion}

\label{sec:Discussion-and-conclusion}

In this paper, we considered photon spheres and images of static and spherical
accretion flows for a class of charged hairy black holes in an EMS model,
where the scalar field is minimally coupled to Einstein's gravity. The hairy
black holes were shown to possess at least one unstable photon spheres.
Intriguingly, we found that the hairy black holes have two unstable and one
stable photon spheres outside the event horizon in a certain parameter regime.
Normally, there is a single unstable photon sphere and no stable photon sphere
outside the event horizon in asymptotically flat black hole spacetime.
Instead, stable photon spheres can appear in horizonless ultra-compact objects
\cite{Cunha:2017qtt} or on the horizons of extreme static black holes
\cite{Tang:2017enb}. A heuristic argument suggests that spacetime possessing
stable photon spheres may be unstable due to the existence of very long-lived
modes \cite{Cardoso:2014sna}. Our results provide an interesting example that
asymptotically flat black holes can have a stable photon sphere outside the
event horizon. Effects of the stable photon sphere on the stability of the
hairy black hole spacetime deserve further study.

On the phenomenological side, one wonders whether imprints of the existence of
more than one unstable photon spheres outside the event horizon can be
observed. To tentatively answer this question, we studied images of an
accretion flow around the hairy black holes in the single- and double-peak
potential cases. When the effective potential has a single peak, there is only
one unstable photon sphere appearing as a bright ring at the edge of the black
hole shadow, which is quite similar to various static black holes considered
in
\cite{Zeng:2020dco,Zeng:2020vsj,Qin:2020xzu,Saurabh:2020zqg,Narayan:2019imo}.
While for the double-peak potential, there exist two unstable photon spheres
of different sizes. Only when the effective potential at the smaller photon
sphere is greater than that at the larger photon sphere, the effects of two
photon spheres come into play. In this case, the most remarkable feature is
that there are two concentric bright rings of different radii surrounding the
shadow. The inner bright ring is identified as the boundary of the shadow, and
the outer one is much brighter and has a considerable size.

Finally, we end with a few comments. For a Schwarzschild black hole, another
spherical accretion model, where the radiating gas radially moves in towards
the black hole, was also considered in \cite{Narayan:2019imo}. The authors
showed that the image of infalling accretion flow is quite similar to that of
static accretion flow except that the shadow region in the infalling model is
significantly darker than that in the static model due to the inward gas
motion. Likewise, it is reasonably expected that, for the hairy black hole
considered in this paper, the image of the infalling accretion flow closely
resembles that of the static accretion flow, except with a darker shadow.
Thus, the static accretion model suffices to illustrate effects of single- and
double-peak structures of the effective potential. On the other hand, real
accretion flows are not spherically symmetric, and hence considering more
realistic accretion models would gain more insights into astrophysical black
hole images released in current and future EHT experiments.

Electric charges of astrophysical black holes are usually neglected since
charged accretion flows can lead to prompt discharging. On the other hand,
magnetically charged black holes are possible exotic astrophysical objects due
to the paucity of magnetic monopoles. Finding astrophysical observations of
such objects can greatly expand our understanding of the universe
\cite{Maldacena:2020skw,Bai:2020spd,Ghosh:2020tdu,Liu:2020vsy,Liu:2020bag}. It is worth noting that
under the electromagnetic duality transformation, the electric hairy black
hole solution $\left(  \ref{eq:metric ansatz}\right)  $ can also describe
magnetically charged black holes \cite{Astefanesei:2019pfq}. Therefore, our
results may shed light on astrophysical observations of magnetic black holes.

For an astrophysical black hole with a nonzero spin, while the image of the
accretion flow is expected to be strongly spin dependent, the size of the
shadow is rather insensitive to the value of the spin. So studying spherical
black holes can capture some key features of astrophysical black hole shadows.
Considering a RN black hole of mass $M$, the shadow radius has a minimum value
of $4M$ in the extremal limit when the black hole is `backlit' from a distant
uniform source or illuminated by the above-discussed spherical accretion flows
\cite{Narayan:2019imo}. Note that other accretion models may lead to a bigger
black hole shadow \cite{Gralla:2019xty}. In contrast, the shadow radius of the
hairy black hole considered in this paper can become zero at the critical
charge (see Fig. \ref{fig:dependenceQ}). The observation of a black hole
shadow with a radius much smaller than $4M$ may indicate modified/alternative
theories of gravity. Our results can provide a viable scenario for this.

\begin{acknowledgments}
We thank Guangzhou Guo for his helpful discussions and suggestions. This work
is supported in part by NSFC (Grant No. 11875196, 11375121, 11947225 and 11005016).
\end{acknowledgments}

\bibliographystyle{unsrturl}

\end{document}